\documentstyle[12pt]{book}
\begin{document}

\sloppy
\raggedbottom

\chapter* 
{A NEW VERSION OF ALGORITHMIC INFORMATION THEORY\footnote
{\it
This material was presented in a series of lectures at the Santa Fe
Institute, the Los Alamos National Laboratory, and the University of
New Mexico, during a one-month visit to the Santa Fe Institute, April
1995.
}
}
\markright
{A New Version of Algorithmic Information Theory}
\addcontentsline{toc}{chapter}
{A new version of algorithmic information theory}

\section*{G. J. Chaitin,
IBM Research Division,
P.~O. Box 704, Yorktown Heights, NY 10598, USA,
chaitin @ watson.ibm.com}
\section*{}

\section*{1. Introduction}

Algorithmic information theory may be viewed as the result of adding
the idea of program-size complexity to recursive function theory.  The
main application of algorithmic information theory is its
information-theoretic incompleteness theorems.  This theory is
concerned with the size of programs, but up to now these have never
been programs that one could actually program out and run on
interesting examples.

I have now figured out how to actually program the algorithms in the
proofs of the all the key information-theoretic incompleteness
theorems in algorithmic information theory.  I have published this
material electronically in a series of detailed reports [1,2,3,4] that
include a lot of software.  These reports are available on the Web at
{\tt http://xyz.lanl.gov} or by sending e-mail to {\tt chao-dyn @
xyz.lanl.gov} with {\tt Subject: get 9506003}, where {\tt yy} is the
year, {\tt mm} is the month, and {\tt nnn} is the number of each
report.

Here I shall give an overview of this material.  I'll present the main
ideas, which are a mixture of abstract mathematics and practical
computer software technology.

In order to program out the algorithms in the proofs of the main
information-theoretic incompleteness theorems, I use a stripped-down
version of pure LISP designed especially for this purpose.  The
interpreter for this LISP was originally written in {\sl
Mathematica\/} [1].  A faster version of this interpreter was then
written in {\sl C\/} [2].  [3] uses a slightly different LISP with an
interpreter that is also written in {\sl C}.  In addition, [1] gives
the {\sl Mathematica\/} code for producing a monster diophantine
equation that exhibits randomness in arithmetic.  This equation is
basically just an interpreter for the LISP used in [1] and [2], dressed
up as a diophantine equation.  I haven't taken the trouble yet to
produce a diophantine equation for the LISP used in [3].

The main idea is this:

In algorithmic information theory, given a self-delimiting universal
Turing machine, one defines a program-size complexity measure from it.
If one picks a different universal Turing machine, one gets a slightly
different complexity measure.

Now I pick a specific machine to base my whole theory on.

My universal Turing machine $U$ produces output in the form of LISP
S-expressions, and $U$'s input is a program in the form of a binary
string.  The machine $U$ starts by reading a LISP S-expression $\pi$
from the start of its program tape.  $U$ reads this S-expression $\pi$
from the tape a character at a time in 7-bit ASCII chunks, until
parentheses balance and $U$ knows that it has read the complete
S-expression $\pi$.  Then $U$ starts to evaluate the prefix $\pi$ that
it has read, with the rest of the program tape $\beta$ available as
binary data to the S-expression $\pi$ in a highly controlled manner.
The S-expression $\pi$ can use a READ-NEXT-BIT pseudo-function to read
individual bits from $U$'s program tape.  $\pi$ cannot look at
$\beta$, the rest of $U$'s program tape, directly.  The READ-NEXT-BIT
function always returns the value 0 or 1, but it cannot return an
end-of-data indication.  If the S-expression $\pi$ attempts to read
beyond the end of the binary data $\beta$, the computation aborts.
This forces $U$'s program $\pi\beta$ to be self-delimiting.  (There is
no punishment for failing to read all of $\beta$.)

The main difference between the pure LISP that I use and ordinary pure
LISP is that there is a mechanism for giving binary data to LISP
S-expressions.  There is also a mechanism that makes it possible for
an S-expression whose evaluation never ends to output an infinite set
of S-expressions (which is all that a formal axiomatic system is).
The heart of an interpreter for ordinary LISP is the recursive
expression evaluator EVAL.  The heart of my interpreter is not EVAL,
it is a mechanism for doing a time-limited evaluation called TRY.

\section*{2. LISP}

The above hints are intended for people who already know LISP.  Now
let me present my LISP from the ground up, using lots of examples.
Then I'll use it to get some incompleteness theorems.

In set theory one constructs the entire universe from the empty set
using the set forming operation.  In LISP, one constructs the entire
universe by making lists starting with atoms.  Lists are ordered, sets
are not.

In this LISP the atoms are the 128 7-bit ASCII characters.  The empty
list, which is also an atom, is {\tt ()}; {\tt (abcd)} is a list of
four atoms; and {\tt ((aa)(bb)(cc))} is a nested list with three
elements, each of which is in turn a list with a repeated atom.  The
union of the set of atoms and the set of lists yields the set of
S-expressions.  S-expressions are the universal LISP substance: both
programs and data are S-expressions.

True and false are {\tt 1} and {\tt 0}.

My LISP is so simple that numbers are not provided.  But it is easy to
program unary or base-two arithmetic using lists of 0's and 1's.

Normally, the result of applying the function $f$ to the arguments
$x$, $y$ and $z$ is written $f(x,y,z)$.  In LISP this is written {\tt
(fxyz)}.

Pure LISP is not an imperative language with side-effects.  Rather one
defines functions and evaluates expressions.  One starts with a number
of primitive functions.

In this LISP all the primitive functions have a fixed number of
arguments.  This makes it possible for there to be a parenthesis-free
meta-notation called M-expressions in addition to the ``official''
S-expression notation.

Now I present the primitive functions using examples.

The primitive function QUOTE yields its unevaluated argument, which is
how one distinguishes data from programs.  For example, the
M-expression {\tt '(abc)} denotes the S-expression {\tt ('(abc))}
whose value is the S-expression {\tt (abc)}.  Without the QUOTE, this
would indicate that the function {\tt a} has to be applied to the
arguments {\tt b} and {\tt c}.

The function HEAD gives the first element of a list.
The M-expression {\tt +'(abc)} yields {\tt a}.

The function TAIL gives the rest of a list.
The M-expression {\tt -'(abc)} yields {\tt (bc)}.

JOIN is the inverse of HEAD and TAIL:
{\tt *'a'(bc)} yields {\tt (abc)}.

The predicate ATOM tells whether its argument is an atom or not:
{\tt .'a} yields {\tt 1} and {\tt .'(abc)} yields {\tt 0}.

The predicate EQUAL compares two S-expressions:
{\tt ='a'b} yields {\tt 0} and {\tt ='(ab)'(ab)} yields {\tt 1}.

The pseudo-function IF-THEN-ELSE has three arguments, and evaluates
its second or third argument depending on its first argument.
{\tt /1'x'y} yields {\tt x} and {\tt /0'x'y} yields {\tt y}.

A function definition is a triple consisting of an ampersand followed
by the list of parameters followed by the function body.
For example, the M-expression \verb|('&(xy)y 'a 'b)| denotes the
S-expression \verb|(('(&(xy)y)) ('a) ('b))| whose value is {\tt b}.
In other words, this is an unnamed function of two parameters
whose value is the second parameter.

Here are two additional pieces of notation, which are extremely
convenient.

The M-expression {\tt :xv e} is an abbreviation for
the M-expression \verb|('&(x)e v)|.
In other words, let {\tt x} have value {\tt v} in expression {\tt e}.

The M-expression {\tt :(fxyz)b e} is an abbreviation for
the M-expression \verb|('&(f)e '&(xyz)b)|.
In other words, let the function
{\tt (fxyz)} be defined to be {\tt b} in expression {\tt e}.

We can now present a complete LISP program.  Let's define list
concatenation recursively and use it to concatenate two lists.
{\tt :(Cxy)/.xy*+x(C-xy) (C'(abc)'(def))} yields {\tt (abcdef)}.
Let's say this in words.  Here is how we define the concatenation of
two lists.  If the first list is empty, then the concatenation is the
second list.  Otherwise take the first element of the first list and
join it to the result of concatenating the rest of the first list to
the second list.

Actually, for efficiency's sake concatenation is provided as a
primitive function \verb$^$.  Also, in another concession to
practicality, there are mechanisms for making function definitions
permanent, so that it is not really necessary to give all function
definitions locally.

There is a mechanism for displaying intermediate results.  This is the
pseudo-function DISPLAY which considered as a pure function is just
the identity function, but which has the side-effect of displaying its
argument.  DISPLAY is written as a comma.  Let's change the definition
of concatenation to see how this works.
{\tt :(Cxy)/.,xy*+x(C-xy) (C'(abc)'(def))} finally yields
{\tt (abcdef)} just as before, but first it displays the intermediate
results {\tt (abc)}, {\tt (bc)}, {\tt (c)}, and {\tt ()}.

The primitive function EVAL allows one to construct an S-expression and
then evaluate it.  EVAL is written as an exclamation mark.
For example,
{\tt !'+'(abc)} yields {\tt a}.

The major difference between this LISP and traditional pure LISP is
the primitive function TRY.  TRY, which is written as a question mark,
has three arguments, $\alpha$, $\beta$ and $\gamma$.  The first,
$\alpha$, is a time bound, given in unary notation, that is, as a list
of 1's.  The second argument $\beta$ is the expression to be
evaluated.  And the third argument $\gamma$ is a list of 0's and 1's
which are made available in a highly controlled manner to the
expression $\beta$ that is being evaluated.  To access the binary data
$\gamma$, the S-expression $\beta$ must use two primitive functions
READ-NEXT-BIT ({\tt @}) and READ-NEXT-S-EXPRESSION (\verb|%|), both of
which are pseudo-functions with no argument.

The value of a TRY is a list whose first element is an error
indication if the TRY failed, or the value of the expression $\beta$
wrapped in parentheses if the TRY succeeded.  The first element is
{\tt ?} if the TRY failed because the time $\alpha$ ran out.  The
first element is {\tt !} if the TRY failed because the binary data
$\gamma$ ran out.  And the rest of the value of a TRY consists of
captured intermediate results that were produced by the expression
$\beta$ using DISPLAY.

My LISP has permissive semantics, so that a TRY can only fail by
running out of time or binary data, not by giving bad arguments to a
primitive function.

Note that one is punished if one runs out of binary data, but not
if one fails to read all the binary data.

EVAL and TRY always start with a fresh environment in which each atom
is bound to itself.  EVAL has no time limit, TRY does.  EVAL inherits
the binary data, if any, in its context, while TRY switches to new
binary data.  EVAL is dangerous; it may run forever or may abort its
user.  TRY is safe; it can only run for a fixed amount of time and it
cannot abort its user.  Using EVAL can eat up one's binary data; using
TRY doesn't touch one's binary data.  DISPLAY's from within EVAL take
place normally; DISPLAY's within TRY are suppressed.

What if one TRY's the concatenation example given above, with the
first arguments {\tt ()} {\tt (1)} {\tt (11)} {\tt (111)} {\tt (1111)}
and {\tt (11111)} and the third argument {\tt ()}? In other words,
let's TRY running the concatenation example with longer and longer
time bounds and no binary data.  Here are the values of TRY that one
gets:
{\tt (?)}
{\tt (?)}
{\tt (?(abc))}
{\tt (?(bc)(abc))}
{\tt (?(c)(bc)(abc))}
and
{\tt (((abcdef))()(c)(bc)(abc))}.

The extra pair of parentheses around the result of the concatenation
makes it possible to distinguish the errors {\tt ?} and {\tt !} from
the valid values {\tt (?)} and {\tt (!)}.

If the first argument of TRY is not a list, then there is no time
limit.  Then one immediately gets {\tt (((abcdef))()(c)(bc)(abc))} as
the value of the TRY.

We actually use a machine-independent time limit $\alpha$ that is a
limit on the maximum interpreter stack depth.  The body of a function
definition and expressions that are EVAL'ed or TRY'ed are evaluated
using a time limit $\alpha' = \alpha-1$ that is one unit less than the
time limit $\alpha$ for the containing expression.

The most delicate thing in the interpreter is getting nested TRY's to
work properly.  First of all, within nested TRY's the most
constraining time limit must apply.  Secondly, if time runs out one
must unwind the interpreter stack back up to the correct TRY, which is
the one that imposed the strongest constraint.

A final primitive function CONVERT-TO-BITS (\verb|#|) converts its
argument from an S-expression into the list of the consecutive bits in
the ASCII for its character string representation.  CONVERT-TO-BITS
and READ-NEXT-S-EXPRESSION are in effect inverse functions.

\section*{3. Complexity}

That concludes the presentation of my LISP.  That's all there is to
it!  It's very simple!  It has to be, if we are going to make it into
a diophantine equation.  But this simple programming language is
powerful enough to code all the proofs of my information-theoretic
incompleteness theorems.  This LISP is powerful enough to program an
interpreter for my universal Turing machine $U$ in a single line of
LISP code.  This LISP can also be used as a very high-level assembler
to produce complicated binary programs for $U$.

In fact, here is how we program $U$.  {\tt (Up)}, the result of
running the universal machine $U$ on the binary program $p$, is
defined to be a TRY with no time limit of EVAL of READ-NEXT-S-EXPRESSION
using $p$ as the associated binary data.  More precisely, {\tt (Up)}
is defined to be the M-expression \verb|++?0'!%p| which is just the
S-expression \verb|(+(+(?0('(!(%)))p)))| which says, read a complete
S-expression from the program tape, then TRY to evaluate it with no
time limit using the rest of the program tape as binary data.

Now that we have $U$, let's use it to measure program-size complexity.
Let's define the complexity $H(x)$ of an arbitrary S-expression $x$ to
be the size in bits of the smallest program for $U$ to compute $x$.
In other words, $H(x)$ is the size in bits of the smallest $p$ such
that $U(p) = x$.

What properties does this complexity measure $H$ have?

Let's start by considering two arbitrary S-expressions $x$ and $y$.
Consider the M-expression \verb|*!% *!% ()|, which is
the S-expression \verb|(*(!(%)) (*(!(%)) ()))|, which is 20 characters
and $7 \times 20 = 140$ bits.  This expression says to read two
S-expressions from the binary data, and then evaluate them and put the
results together into a list.  If we concatenate this 140-bit prefix
with a minimum-size program for $U$ to calculate $x$ and a
minimum-size program for $U$ to calculate $y$, we get a program for
$U$ to calculate the pair $(x y)$ that is precisely $140 + H(x) +
H(y)$ bits long.  Thus we see in a very concrete manner that the joint
complexity $H(x,y)$ is bounded by the sum of the individual
complexities $H(x)$ and $H(y)$ plus the constant $c = 140$: $H(x,y)
\le H(x) + H(y) + 140$.

Now consider a binary string $x$ that is $|x|$ bits long.  It's easy
to see that $H(x) \le 2|x| + 471$ and $H(x) \le |x| + H(|x|) + 1148$.

Why?

To prove that $H(x) \le 2|x| + 471$ one programs a 469-bit prefix
$\pi$ so that $U(\pi \, 00 \, 11 \, 00 \, 11 \, 01) = 0 1 0 1$.  In
other words, the prefix $\pi$ makes $U$ read two bits at a time from its
program tape as long as they are equal.  $U$ stops as soon as two
unequal bits are found.

And to prove that $H(x) \le |x| + H(|x|) + 1148$ one programs a
1148-bit prefix $\pi$ so that $U(\pi \beta x ) = x$ if $\beta$ is
a minimum-size program for $U$ to compute the base-two notation for
the size of the bit string $x$.  Here is how this works: first the
prefix $\pi$ makes $U$ run the program that follows $\pi$ on the
program tape to determine how many bits there are, then $\pi$ makes
$U$ read that many bits from the program tape and stop.

For more details, see the program {\tt univ.lisp} in [2].

Next we show how to compute the halting probability $\Omega$ of $U$ in
the limit from below.  TRY running $U$ for time $N$ on each $N$-bit
program.  This gives a monotone increasing sequence $W_N$ of lower
bounds on $\Omega$: $W_N$ is (the number of $N$-bit programs that halt
on $U$ within time $N$) divided by $2^N$.

Here is a simple way to calculate $W_N$.  Given $N$, first generate
the list of all $N$-bit strings.  Then use TRY to select the subset of
this list of programs that halt within time $N$ when run on $U$.  Then
count the size of the list of programs that halt, and develop this
count in base-two notation.  Finally pad the base-two count on the
left with enough 0's to make an $N$-bit string, and prepend ``0.''.
By using a slightly more sophisticated counting technique, one can
avoid having to generate these two enormous lists of programs.  That's
a good idea, because my fast {\sl C\/} interpreter does not have
garbage collection.  In this manner I was able to compute $W_{22}$ in
about an hour on a 512-megabyte IBM RS/6000 workstation.

For more details, see the program {\tt omega.lisp} in [1] and [2].

Now we show that the bits of the halting probability $\Omega$ are
irreducibly complex.  More precisely, let $\Omega_N$ be the first $N$
bits of the base-two representation of the real number $\Omega$.  We
shall show that $H(\Omega_N) > N - 4431$.

Why?

The reason is that knowing $\Omega_N$ enables one to solve the halting
problem for all $N$-bit programs.  More precisely, there is a prefix
$\pi$ that is 4431 bits long that when concatenated to a minimum-size
program $\omega$ for $U$ to compute $\Omega_N$ does the following.
First $\pi$ runs $\omega$ to compute $\Omega_N$.  Then $\pi$
calculates $W_T$ for larger and larger values of $T$ until the first
$N$ bits of $W_T$ are okay, that is, agree with $\Omega_N$.  At this
point $\pi$ knows that any $N$-bit program for $U$ that halts must
halt within time $T$.  So $\pi$ can give as its final result the list
$B_N$ of the values of all $N$-bit programs for $U$ that halt.  $B_N$
includes every S-expression with program-size complexity $\le N$.
Hence this big list $B_N$ must itself have program-size complexity
greater than $N$.  Thus $N < H(B_N) \le |\pi\omega| = 4431 +
H(\Omega_N)$, which was to be proved.

For more details, see the program {\tt omega2.lisp} in [2].

Now let's get an incompleteness result.  We show that a formal
axiomatic system of complexity $N$ cannot enable us to determine more
than $N + 4431 + 3150$ bits of the binary representation of the
halting probability $\Omega$.

Why?  It's just a version of the Berry paradox discussed in [5],
which deals with ``the first positive integer that cannot be named
in less than a billion words.''

This time we employ a prefix $\pi$ that is 3150 bits long and contains
within itself the binary constant for the sum of 3150 and 4431.  We
concatenate this prefix $\pi$ to the formal axiomatic system $\alpha$,
which is an unending minimum-size program for $U$ to DISPLAY
S-expressions of the form {\tt (110X0XXX10)} standing for partial
determinations of the bits of $\Omega$.  At any given time the prefix
$\pi$ has read $\rho$ bits of $\alpha$ as it TRY's to DISPLAY more and
more theorems of the formal axiomatic system $\alpha$.  This continues
until $\pi$ is able to determine more than $4431 + 3150 + \rho$ bits
of $\Omega$ using $\rho$ bits of $\alpha$.  At that point the missing
{\tt X} bits of $\Omega$ are read by $\pi$ from the program tape at a
cost of exactly one bit added for each missing bit.  The result is a
program for $U$ that is exactly $3150 + \rho + {} $(the number of
missing bits) bits long that calculates a complete initial segment of
the binary expansion of $\Omega$ that is more than 4431 bits longer
than it is.  But this contradicts the previously established fact that
$H(\Omega_N) > N - 4431$.

For more details, see the program {\tt godel3.lisp} in [2].

There is also a more sophisticated M-expression version of our
universal Turing machine $U$.  It is used in [3].  This new and
improved version of $U$ reads an M-expression prefix from the
beginning of the program tape, not an S-expression prefix.  Then, as
before, $U$ runs this M-expression with the rest of the program tape
as binary data.  To change $U$ to work this way only requires that two
primitive functions, READ-NEXT-S-EXPRESSION and CONVERT-TO-BITS, be
changed to work with M-expressions instead of S-expressions.  And in
[3] we also use a more aggressive kind of M-expression, in which
defined functions as well as primitive functions have their
parentheses omitted.  Thus in [3] our concatenation example simplifies
to
{\tt :(Cxy)/.xy*+xC-xy C'(abc)'(def)}.
The constants in our previous results, which come from [2], shrink
substantially in [3].  Now we get
$H(x,y) \le H(x) + H(y) + 56$,
$H(x) \le 2|x| + 142$,
$H(x) \le |x| + H(|x|) + 441$,
$H(\Omega_N) > N - 1883$,
and a formal axiomatic system of complexity $N$ cannot enable us to
determine more than $N+2933$ bits of the binary representation of the
halting probability $\Omega$.

\section*{4. Arithmetic}

The halting probability $\Omega$ shows that some mathematical
questions are irreducible, that is, have the property that essentially
the only way to prove them is to add them as new axioms.  The bits of
the $\Omega$ are irreducible mathematical facts.  But can we find
irreducible mathematical facts in elementary number theory?  The
answer is yes, and I can even explicitly exhibit arithmetical versions
of the bits of $\Omega$.  I do this by using {\sl Mathematica\/} to
convert an interpreter for the LISP used in [1] and [2] into an
enormous diophantine equation.  (I haven't done this work yet with the
LISP used in [3].)

Here is an outline of how to get a diophantine equation that can be
used as a LISP interpreter.  (For the programming details and the {\sl
Mathematica\/} software, see [1].  For a detailed explanation, see
Chapter 2 of [6].)

The first step is to write abstract register machine code for a LISP
interpreter that works one character at a time on the reversed
character strings for S-expressions.  It's hard work to debug such
low-level code!  In order to speed up the job, I use a {\sl
Mathematica\/} program to compile the register machine code into {\sl
C.} Then I use another {\sl Mathematica\/} program to compile the
register machine code into an exponential diophantine equation.
(``Exponential'' just means that unknowns can occur in exponents.)
This equation $L(k,x_1,x_2,\ldots) = R(k,x_1,x_2,\ldots)$ is several
hundred pages long and has tens of thousands of unknowns.  Let the
positive integer $k$ be the ASCII representation of the reversal of
the character string representation of the S-expression $K$.  Then
$L(k,x_1,x_2,\ldots) = R(k,x_1,x_2,\ldots)$ is constructed so that it
has precisely one solution in positive integers $x_1,x_2,\ldots$ if
evaluation of the LISP S-expression $K$ halts, and this equation is
constructed so that it has no solution in positive integers
$x_1,x_2,\ldots$ if evaluation of the LISP S-expression $K$ fails to
halt.

Plug into this monster equation an S-expression $K$ that loops forever
if the $k$th bit of $W_n$ is a 0 and that halts if the $k$th bit of
$W_n$ is a 1.  Then $L(k,n,x_1,x_2,\ldots) =
R(k,n,x_1,x_2,\ldots)$ has finitely many solutions in positive
integers $n,x_1,x_2,\ldots$ if the $k$th bit of $\Omega$ is a 0,
and this equation has infinitely many solutions in positive integers
$n,x_1,x_2,\ldots$ if the $k$th bit of $\Omega$ is a 1.

Thus whether $L(k,\ldots) = R(k,\ldots)$ has finitely or infinitely
many positive integer solutions is so delicately balanced that it is
completely accidental whether it goes one way or the other!  In other
words, this is a completely accidental mathematical fact, it is a
mathematical fact that is true for no reason, and therefore escapes
the power of mathematical reasoning.

What are we to make of this incompleteness result?  I have discussed
this at length in my reductionism paper [7].  Briefly, it makes me
ask, is mathematics quasi-empirical?  In other words, should one
perhaps add new axioms to elementary number theory based on the
results of computer experiments?!

To conclude, I think that the incompleteness results of algorithmic
information theory seem much more real and concrete now that the
programming details have been worked out and all this software is
available.

\section*{References}

\begin{itemize}

\item[{[1]}]
G. J. Chaitin,
``The limits of mathematics---Course outline \& software,''
chao-dyn/9312006,
{\it IBM Research Report RC-19324,} 127 pp., December 1993.

\item[{[2]}]
G. J. Chaitin,
``The limits of mathematics,''
chao-dyn/9407003,
{\it IBM Research Report RC-19646,} July 1994, 270 pp.

\item[{[3]}]
G. J. Chaitin,
``The limits of mathematics IV,''
chao-dyn/9407009,
{\it IBM Research Report RC-19671,} July 1994, 231 pp.

\item[{[4]}]
G. J. Chaitin,
``The limits of mathematics (Extended abstract),''
chao-dyn/9407010,
{\it IBM Research Report RC-19672,} July 1994, 7 pp.

\item[{[5]}]
G. J. Chaitin,
``The Berry paradox,''
{\it Complexity\/} 1 (1995), pp.\ ???--???.

\item[{[6]}]
G. J. Chaitin,
{\it Algorithmic Information Theory,}
Cambridge University Press, 1987.

\item[{[7]}]
G. J. Chaitin,
``Randomness in arithmetic and the decline and fall of reductionism in
pure mathematics,''
in J. Cornwell, {\it Nature's Imagination,}
Oxford University Press, 1995, pp.\ 27--44.

\end{itemize}

\end{document}